# Unveiling the Potential of Graph Neural Networks in SME Credit Risk Assessment

1st Bingyao Liu
University of California, Irvine
Irvine, USA

2nd Iris Li
New York University
New York, USA

3rd Jianhua Yao
Trine University
Phoenix, USA

4th Yuan Chen
Rice University
Houston, USA

5th Guanming Huang
University of Chicago
Chicago, USA

6th Jiajing Wang*
Columbia University
New York, USA

*Abstract*—This paper takes the graph neural network as the technical framework, integrates the intrinsic connections between enterprise financial indicators, and proposes a model for enterprise credit risk assessment. The main research work includes: Firstly, based on the experience of predecessors, we selected 29 enterprise financial data indicators, abstracted each indicator as a vertex, deeply analyzed the relationships between the indicators, constructed a similarity matrix of indicators, and used the maximum spanning tree algorithm to achieve the graph structure mapping of enterprises; secondly, in the representation learning phase of the mapped graph, a graph neural network model was built to obtain its embedded representation. The feature vector of each node was expanded to 32 dimensions, and three GraphSAGE operations were performed on the graph, with the results pooled using the Pool operation, and the final output of three feature vectors was averaged to obtain the graph's embedded representation; finally, a classifier was constructed using a two-layer fully connected network to complete the prediction task. Experimental results on real enterprise data show that the model proposed in this paper can well complete the multi-level credit level estimation of enterprises. Furthermore, the tree-structured graph mapping deeply portrays the intrinsic connections of various indicator data of the company, and according to the ROC and other evaluation criteria, the model's classification effect is significant and has good "robustness".

## I. INTRODUCTION

Enterprises, as the main participants in the market economy, especially small and medium-sized enterprises (SMEs), are increasingly important in the national economy, absorbing the largest group of employees and being an essential component of our country's economic development. The issue of corporate credit risk has always been a focus of attention in academic and financial circles. As early as in the "Basel Accord," the prediction of corporate credit risk was elaborated and regarded as an indispensable part of financial activities. Corporate credit risk is an important part of financial risk. With the development of the commodity economy and the growth of enterprises, enterprises are no longer isolated individuals in economic activities but have become an inseparable whole with multiple enterprises in various economic activities, which has intensified the contagiousness of corporate credit risk, making the spread of financial risk more complex and diverse. If corporate credit risk is not handled properly, it is very likely to lead to systemic financial risk.

For this reason, how to effectively assess the credit risk of SMEs has attracted the attention of researchers.

Traditional assessment methods rely on expert subjective judgment. Faced with such nonlinear, non-stationary, and volatile time series data [1], they are often time-consuming and labor-intensive, and cannot quickly and timely provide credit ratings for various enterprises. Nowadays, with the rapid development of computer technology, machine learning has shown excellent results in many fields[2-4]. Classical algorithms such as KNN[5], SVM [6], NB [7], LSTM [8], neural networks, etc., have been widely used in the financial field. In response to these challenges, there has been a significant shift towards leveraging advanced computational technologies. Recent advancements in machine learning, particularly the emergence of graph neural networks (GNN), offer promising new directions for credit risk assessment. Graph neural networks are adept at modeling complex patterns and relationships within data, making them particularly suited for analyzing the intricate web of financial indicators within enterprises. By abstracting each financial indicator as a vertex and the relationships between them as weighted edges, GNNs facilitate a more nuanced understanding of corporate structures and their inherent risk profiles.

This paper proposes a novel GNN-based model that integrates and analyzes the intrinsic connections between enterprise financial indicators, offering a sophisticated approach to SME credit risk assessment. By constructing a detailed similarity matrix and utilizing graph-structural techniques, the model captures the multi-dimensional nature of financial data, thus enabling a more accurate prediction of credit risk levels. Our approach not only enhances the predictive accuracy but also contributes to the broader understanding of financial contagion mechanisms, thereby supporting more informed and effective financial risk management strategies.

## II. RELATED WORK

In recent years, the application of Graph Neural Networks (GNNs) in financial and risk prediction tasks has gained significant attention. GNN-based models have been shown to be particularly effective in financial fraud detection, lending decisions, and credit risk analysis. Cheng et al. proposed an advanced GNN-CL model for financial fraud detection, emphasizing the importance of graph-structured data in identifying fraudulent behavior in financial systems [9]. This study demonstrates the capability of GNNs to model the complex relationships between financial indicators, which is closely aligned with the focus of this paper on SME credit risk.

Similarly, Wei et al. examined the role of feature extraction and model optimization in deep learning, specifically targeting stock market predictions. Their approach highlighted the predictive power of deep learning models in financial forecasting, providing insights into how deep learning can be applied to dynamic financial datasets [10]. This work complements the methods used in this study, where enterprise financial data is analyzed using a GNN-based approach to assess credit risk.

Beyond direct applications to financial systems, research in other domains such as image recognition and natural language processing offers valuable methodologies that can be adapted for financial analysis. Zhong et al. compared various Generative Adversarial Networks (GANs) and traditional image recognition methods, emphasizing the role of deep learning in achieving higher accuracy and efficiency [11]. This study's exploration of advanced deep learning architectures can inform the optimization of GNN models used in financial systems. In a similar vein, Mei et al. investigated the efficiency optimization of large-scale language models, focusing on the performance of deep learning in natural language processing tasks [12]. The optimization strategies presented in these works could be adapted to enhance the efficiency of GNN models in credit risk assessment.

Furthermore, the development of enhanced deep learning architectures such as the Encoder-Decoder network presented by Gao et al. has demonstrated the potential to reduce information loss in complex tasks like semantic segmentation [13]. While this research focuses on image segmentation, the underlying principles of minimizing data loss are relevant for ensuring the integrity of financial data when processed by GNNs. Similarly, Bo et al.'s work on attention mechanisms in text mining and machine translation offers insights into improving context modeling, which can be crucial when dealing with enterprise financial data in a graph-based system [14].

Finally, the computational optimization techniques for deep learning models in different contexts, as explored by Chen et al. and Zheng et al., offer potential enhancements that can be applied to the SME credit risk model presented in this paper. Chen et al. worked on optimizing computations for multivariate polynomial matrices, which could be useful in optimizing the performance of GNN models used for financial predictions [15]. Similarly, Zheng et al. introduced adaptive friction in deep learning optimizers, a technique that can improve the convergence speed and accuracy of deep learning models, potentially benefiting GNN-based credit risk models [16].

## III. METHOD

### A. Rating Method

This paper builds upon previous research on corporate credit risk and comprehensively considers five aspects of a company's financial health: debt-paying ability, profitability, operational capacity, cash generation ability, and development capacity. From these aspects, a total of 29 indicators were selected, including total assets, cash and cash equivalents, net assets, total liabilities, interest-bearing debt, net debt, cash flow from operating activities, cash flow from investing activities, cash flow from financing activities, main business revenue, main business profit, EBITDA, net profit, main business profit margin, main business revenue growth rate, total asset return rate, return on net assets, EBITDA/total operating revenue, operating cash flow/EBITDA, current ratio, quick ratio, inventory turnover rate, asset-liability ratio, short-term debt/total debt, interest-bearing debt/total capital employed, cash ratio, cash and cash equivalents/total debt, interest coverage ratio, and EBITDA/interest-bearing debt. In consideration of the situation of SMEs, enterprises with total asset scale in the top 20% are excluded.

The credit rating method used in this paper is the Z-SCORE Rating [17], which quantifies the likelihood of credit risk default by sorting the implied yield rates of mainstream existing bonds in the market based on past transaction information, such as publicly issued medium-term notes, short-term financing bills, corporate bonds, and enterprise bond data, etc.

The Z-SCORE Rating divides the credit into 1-10 levels, with the smaller the number, the higher the level, and it is currently the most recognized credit assessment method in the domestic market for several reasons:

For subjects with a Z-SCORE Rating above 6, there have been very few historical defaults, and subjects above 7 are relatively safe; The samples of Z-SCORE Rating are normally distributed when assessing the credit level of the subject, which is more reasonable; The distance from the first downgrade date to the default date of Z-SCORE Rating is mainly distributed over 1 year, while the external agency rating downgrades are mainly distributed within 3 months, showing stronger forward-looking nature of the rating adjustments.

## B. Overview of Model Architecture

The model proposed in this paper is mainly divided into three parts: The first step is the graph mapping of enterprises, the second step is the embedding representation of the graph, and the third step is the MLP classifier.

Graph mapping of enterprises:

a) Enterprise data input: Obtain standardized annual report data for each enterprise, where each indicator corresponds to a vector, creating an indicator matrix as the input data for the enterprise;

b) Vertex similarity matrix: Treat each indicator (each column of the matrix) as a vertex, define similarity using the cosine value, and traverse all vertex pairs to obtain the similarity matrix;

c) Construction of graph mapping: Based on the vertex similarity matrix, construct a maximum weighted supporting graph, mapping enterprises as the smallest connected graph with the maximum similarity, denoted as Corp_Tree.

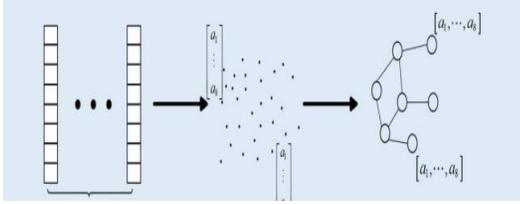

Figure 1. Graph mapping of enterprises

Embedding representation of the graph:

a) First, we expanded the feature vector of each node to 32 dimensions to capture a comprehensive representation of enterprise financial data, which involves complex and interrelated indicators. The choice of 32 dimensions was determined through experimentation, where it balanced model complexity and computational efficiency effectively. This dimensionality ensures that the neural network can learn significant patterns without being overfitted to the noise in the data. Further, this dimensionality was optimized to capture both local and global patterns in the graph structure, enhancing the model's predictive accuracy and robustness.

b) Then perform three GraphSAGE operations on the graph, and the results are pooled using the Pool operation;

c) Finally, average the 3 vectors MEAN to obtain the graph embedding vector.

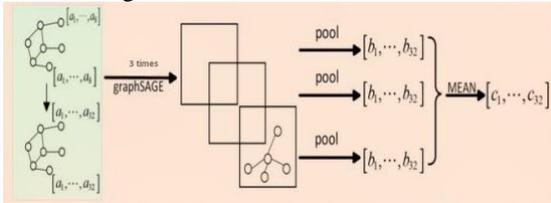

Figure 2. Embedding representation of the graph

MLP classifier: Finally, combine the neural network classifier to complete the prediction task.

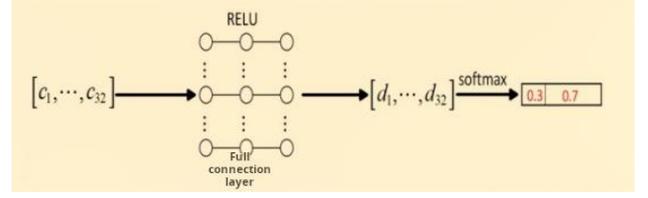

Figure 3. MLP classifier

## C. Details of Model Architecture

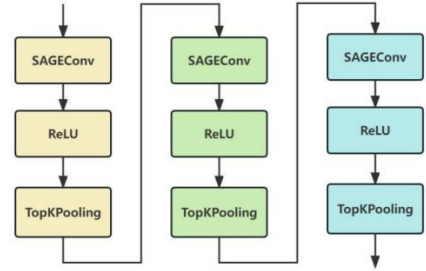

Figure 4. Convolution Process Flowchart

This paper utilizes the greedy algorithm within the maximum spanning tree algorithm to obtain a tree with the highest similarity. The core idea of the algorithm is to sort the edges by their weights, starting with the edge with the largest weight and adding it to the spanning tree. However, each time an edge is selected, it must be checked whether the edge, when added to the tree, forms a cycle with the edges already in the tree. If a cycle is formed, the edge is discarded, and this process continues until m edges are selected.

The specific process of the algorithm is as follows:
Input: A company subgraph $G$ with a node set $V = \{x_B, \ldots \ldots x_{n-l}\}$, where n is the number of vertices, and an edge set $E$, $T=(U, TE)$ is the maximum spanning tree of $G$, where $U$ is the set of vertices in $T$, and $TE$ is the set of edges in T, initially an empty set.

a) Sort all edges by weight in descending order and add the edge with the largest weight to the set T.

b) Compare the remaining edges and select the edge with the largest current weight to add to the spanning tree.

c) Repeat the above steps, discarding any edge that would form a cycle with T when added, until a maximum spanning tree with n - 1 edges is formed.

This paper employs the TopKPooling method for model pooling, as illustrated in Figure 5. TopKPooling was selected for its ability to focus on the most informative nodes, enhancing model performance by preserving critical graph features while reducing dimensionality. It outperformed average and max pooling in our tests by maintaining essential node features without diluting distinctive cues.

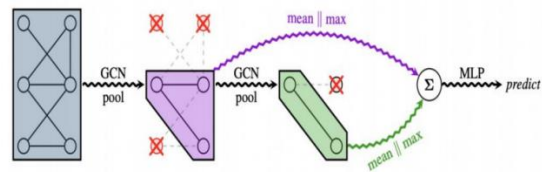

Figure 5. TopKPooling

Finally, a neural network classifier is combined to complete the prediction task with the following steps:

The embedding vectors of the graph generated in the second step are input into a two-layer fully connected neural network. The hidden layer is activated using the ReLU (Rectified Linear Unit) function. The final layer employs the Softmax function to output the classification vector.

## IV. EXPERIMENT

### A. Experiment Settings

The data used in this experiment is sourced from the WRDS COMPUSTAT database [18], which represents real and existing information. The time span of this data set ranges from 2014 to 2021, a total of 8 years. The dataset includes a total of 5739 companies, with an initial data volume of 49584 entries. The dataset is handled through the Linked Data methodology, which consolidates various data formats, a crucial aspect in academic studies[19]. This organized technique facilitates the cross-referencing of data, boosting the interoperation among different datasets. This functionality proves particularly useful in domains like machine learning and artificial intelligence, where the caliber of data is vital for the effective training of models and securing precise outcomes. The rating scores are based on the Z-SCORE Rating system, with a total of 30790 rated entries. Model training for 3-category, 5-category, and 8-category classifications is conducted on two types of graph mapping structures: Corp_Tree and Corp_Tree+.

### B. Experiment Results Analysis

In the experiment, the Corp_Tree tree structure was used for 3-category classification with a learning rate update mode of warm restarts. The changes in the loss function and the ROC curve are shown in the following figures. At this point, we stop updating the parameters, and the overall accuracy of the test set reaches 72%.

According to the ROC curve, we can see that the classifier on the tree structure has a significant effect, and the classification effect for the best credit level (Category 0) is the best, indicating that the model can generally complete the classification task well.

Under the same settings, we performed the 3-category task on the Corp_Tree+ structure (which adds the top 10 remaining high-weight edges to this structure, enhancing connectivity and capturing more nuanced relationships, which improved classification accuracy in our model), and the loss function and ROC curve are shown in the following figures.

On the Corp_Tree tree structure for 5-category classification, the same learning rate update mode of warm restarts was used, with the data being consistent with the previous data. The changes in the loss function and the ROC curve are shown in the following figures.

From Figures 6-7, it can be seen that the 5-category results are generally similar to the 3-category. In terms of the ROC curve, the overall effect of the classifier established by the model is still good, and the classification effect for the best credit level (Category 0) remains optimal. However, compared to the 3-category, there is a certain weakening in the classification ability, and the micro-average, which measures the overall classification, has a noticeable decrease. This is because when the total amount of data remains unchanged, and the number of categories increases, the number of samples corresponding to each category decreases, which to some extent affects the performance of the model.

However, we also noticed that when performing the 8-category classification on the Corp_Tree tree structure, although the effect has weakened compared to the 3-category situation, the overall classifier still has good significance, and the effect of the highest credit rating category is still optimal. Overall, the model did not experience a serious decline in performance due to the further reduction in samples corresponding to each category.

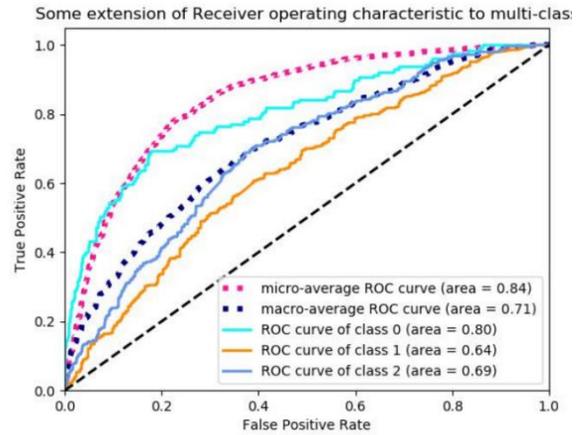

Figure 6. Corp_Tree+ Structure 3-Category Operation Results

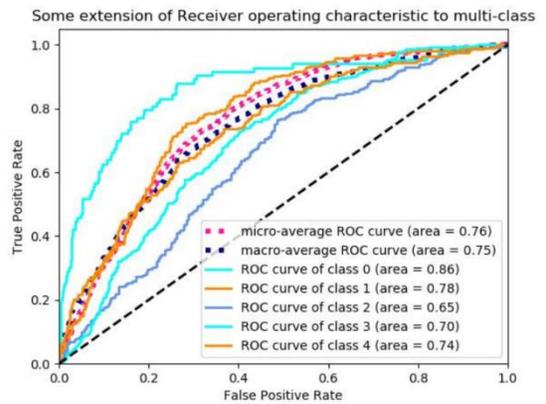

Figure 7. Corp_Tree+ Structure 5-Category Operation Results

## V. CONCLUSIONS

This paper presents a cutting-edge approach to SME credit risk assessment by leveraging Graph Neural Networks (GNN), which effectively capture and model the intricate relationships among various enterprise financial indicators. By abstracting each financial indicator as a vertex and establishing weighted edges based on their similarities, the proposed methodology constructs a comprehensive graph representation of each enterprise's financial landscape. This graph-based framework enables advanced representation learning through GNN, allowing the model to discern both local and global patterns within the data. The experimental results, derived from real-world enterprise datasets, demonstrate that the proposed model not only achieves high accuracy in multi-level credit risk classification but also exhibits robust performance across multiple evaluation metrics, including ROC curves and other key indicators of classification effectiveness [20]. Moreover, the tree-structured graph mapping provides deeper insights into the intrinsic connections and potential contagion effects among corporate financial indicators, thereby enhancing the understanding of financial risk dynamics. Compared to traditional machine learning techniques, which often rely on isolated feature analysis and subjective expert judgment, the GNN-based approach offers a more holistic and data-driven solution, improving predictive accuracy and reducing the time and labor associated with credit risk assessment. Additionally, this research highlights the scalability and adaptability of GNNs in handling complex and high-dimensional financial data, making it a valuable tool for financial institutions aiming to mitigate credit risk and support sustainable growth among SMEs. Future work could explore the integration of temporal dynamics and external economic factors into the graph model, further refining risk predictions and enhancing the model's applicability in diverse financial environments. Ultimately, the proposed GNN-based methodology not only advances the field of financial risk assessment but also contributes to the broader objective of fostering economic stability and growth by providing more reliable and nuanced credit evaluations for small and medium-sized enterprises.